\documentclass[%
 reprint,
 amsmath,amssymb,
 aps,
]{revtex4-2}

\usepackage{graphicx}% Include figure files
\usepackage{dcolumn}% Align table columns on decimal point
\usepackage{soul}
\usepackage{bm}% bold math
\usepackage[colorlinks,linkcolor=blue,citecolor=blue,urlcolor=blue]{hyperref}
\begin{document}

%\title{The field-induced insulating dome of graphite under pressure }

\title{Tuning the BCS-BEC crossover of electron-hole pairing with pressure}

\author{Yuhao Ye$^{1}$, Jinhua Wang$^{1}$, Pan Nie$^{1}$, Huakun Zuo$^{1}$, Xiaokang Li$^{1}$, Kamran Behnia$^{2}$, Zengwei Zhu$^{1,*}$, and Beno\^{\i}t Fauqu\'e$^{3}$}

\affiliation{(1) Wuhan National High Magnetic Field Center and School of Physics, Huazhong University of Science and Technology,  Wuhan  430074, China\\
	(2) Laboratoire de Physique et d'Etude des Mat\'{e}riaux (CNRS)\\ ESPCI Paris, PSL Research University, 75005 Paris, France\\
	(3) JEIP, USR 3573 CNRS, Coll\`ege de France, PSL University, 11, place Marcelin Berthelot, 75231 Paris Cedex 05, France\\
}

\date{\today}

\begin{abstract}
	
	In graphite, a moderate magnetic field confines electrons and holes into their lowest Landau levels. In the extreme quantum limit, two insulating states with a dome-like field dependence of the their critical temperatures are induced by the magnetic field. Here, we study the evolution of the first dome (below 60 T) under hydrostatic pressure up to 1.7 GPa. With increasing pressure, the field-temperature phase boundary shifts towards higher magnetic fields, yet the maximum  critical temperature remains unchanged. According to our fermiology data, pressure amplifies the density and the effective mass of hole-like and electron-like carriers. Thanks to this information, we verify the persistent relevance of the BCS relation between the critical temperature and the density of states in the weak-coupling boundary of the dome. In contrast, the strong-coupling summit of the dome does not show any detectable change with pressure. We argue that this is because the out-of-plane BCS coherence length approaches the interplane distance that shows little change with pressure. Thus, the BCS-BEC crossover is tunable by magnetic field and pressure, but with a locked summit. 
\end{abstract}
\maketitle

\section{INTRODUCTION}
\begin{figure*}
	\centering
	\includegraphics[width=18cm]{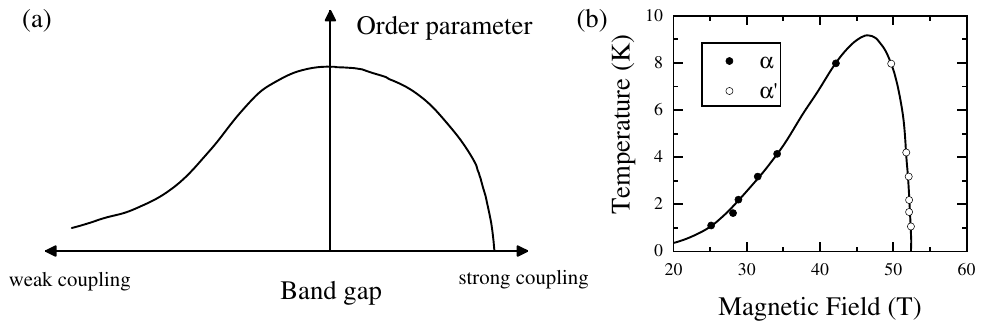}
	\caption{ \textbf{Comparing a theoretical and an experimental phase diagram} (a) The theoretical phase diagram for an excitonic insulator showing the evolution of the ordering energy as a function of the band gap \cite{kozlov1965metal,Jerome1967}. The order parameter is strongest when the band gap is zero. Note the contrast between the evolution of the order parameter on the two sides of the dome. (b) The experimental phase diagram of graphite at high magnetic field \cite{yaguchi1998,Hiroshi2009}. The insulating state resides inside a dome in the (field, temperature) plane. A second dome (starting at $\approx 53$ T and ending at $\approx$ 70 T \cite{Fauque2013}) is not shown. Note the contrast between the gradual rise of the critical temperature to the summit of the dome and its abrupt drop afterwards. }
	\label{figure0}
\end{figure*}

In 1961, Mott made the observation that Coulomb attraction between electrons and holes of a semi-metal can form bound pairs known as excitons \cite{Mott1961}.  Knox then proposed that a sufficiently large exciton binding energy would lead to an insulating state, quite distinct from an ordinary band insulator \cite{Knox1963}. Later, Keldysh and Kozlov \cite{keldysh1967} remarked that if the carriers are sufficiently light and not too dilute, the bosonic excitons would have a sizeable Bose-Einstein condensation (BEC) temperature. Starting from these two postulates, the early research on excitonic insulators proposed that this state of matter should be sought near a semimetal to semiconductor transition and produced a phase diagram, which we reproduce in  Fig. \ref{figure0}a (See figure 3 in \cite{kozlov1965metal}, figure 1 in \cite{Jerome1967} and figure 3 in \cite{Abrikosov1974}). In 1985, Nozi\`eres and Schmitt-Rink \cite{Nozieres1985} demonstrated that the transition between the strong-coupling limit (the BEC of composite bosons, either excitons or Cooper pairs) to the weak-coupling limit (the Bardeen–Cooper–Schrieffer or BCS) is smooth. The latter corresponds to the long tail on the left hand side of the excitonic dome in Fig. \ref{figure0}a.   

Graphite, a semimetal with an equal density of electrons and holes ($n=p\approx3\times10^{18} \rm{cm}^{-3}$ \cite{Brandt1988}), suffers a phase transition at high magnetic field \cite{Fauque2016}, which has been under exploration during four decades \cite{Yoshioka1981,yaguchi1998,Hiroshi2009,Fauque2011,Akiba2015,Fauque2013,LeBoeuf2017,Zhu2019,Wang2020,Marcenat2021}. The experimental discovery in 1981 \cite{Tanuma1981} led to an immediate theoretical identification of this state \cite{Yoshioka1981} as a charge density wave (CDW). Indeed, confining all carriers to their lowest Landau level opens the way to a nesting instability. This is the case of graphite in presence of a magnetic field exceeding 7.4 T \cite{Zhu2010,Schneider2012}. In 1998, Yaguchi and Singleton discovered that the field-induced state abruptly ends at 53 T \cite{yaguchi1998} (see Fig. \ref{figure0}b). In 2013, Fauqu\'e \textit{et al.} found that the first dome is followed by a second dome\cite{Fauque2013} and that the $c$-axis resistance shows an activated behavior in both domes. These observations challenged the CDW scenario. In 2017, Zhu \textit{et al.} highlighted the similarity between the experimental (Fig. \ref{figure0}b) phase diagram of graphite and the theoretical (Fig. \ref{figure0}a) phase diagram of an excitonic insulator\cite{Zhu2017}. The accumulated experimental evidence since then indicates that while the transition can be described by a BCS picture of electron-hole pairing at low field \cite{Hiroshi2009,LeBoeuf2017,Marcenat2021}, the summit of the dome corresponds to the temperature at which the thermal wavelength and the interbosonic distance match \cite{Wang2020}, as expected for a BEC transition \cite{Silvera1997}.  

Besides excitons \cite{Wilson2021,Wang2019exciton,Li2017_grahene,Butov2007,Liu2017,Kogar2017,Lu2017},  BEC has been reported for other bosonic systems, like photons \cite{Klaers2010,Bloch2022}, microcavity polaritons \cite{Deng2010,Kasprzak2006}, and magnons \cite{Zapf2014,Demokritov2006,Giamarchi2008}. On the other hand, the BCS-BEC crossover \cite{CHEN20051,Strinati2018}, which requires tuning either distance between the bosons or the  BCS correlation length, has been mainly studied in ultracold Fermi gases, thanks to the Feshbach resonance \cite{Jochim2003BCSBEC,Greiner2003BCSBEC,Zwierlein2003BCSBEC,Regal2004BCSBEC,Kinast2004BCSBEC,Bourdel2004BCSBEC,Zwierlein2004BCSBEC,Bartenstein2004BCSBEC}. The possibility of the existence of BEC-BCS crossover in superconductors has been proposed for cuprates \cite{CHEN20051}, organic superconductors \cite{Suzuki2022,Mckenzie1997Science}, iron-based superconductors \cite{Kang2020FeSe,Faeth2021}, gate-controlled two-dimensional superconducting devices \cite{Saito2016NRM,Nakagawa2021Science}, interfacial superconductors \cite{Richter2013nature,Bozovic2020NP}, magic-angle twisted superconducting bilayer \cite{Oh2021Nature,Cao2018Nature} trilayer graphene \cite{Kim2022nature,Park2021nature}, and magnetoexcitonic condensates in heterostructure superconducting graphene \cite{Liu2022Science}.

%which represent an indispensable tool for fine-tuning the interactions among atoms in ultracold quantum gases.
%Another issue is the ability to tune the interaction strength between bosons of BEC. 
%For instance, through numerous experimental applications this method,  some important breakthroughs in the field have been facilitated\cite{Bloch2008,Chin2010,Zhang2020,Pethick2008}, leading to two new states of matter liquid-like self-bound droplets and supersolid crystals formed from these droplets etc.
%The Feshbach resonances can also control a variety of physical properties of BEC of cold atoms, such as tuning scattering length, density, critical temperature, dipole moments and excitation spectra\cite{Bloch2008,Chin2010,Pethick2008}.

Here, we present a systematic study of the evolution of the phase diagram of graphite and its Fermi surface by measuring the magnetoresistance for $H\parallel c$-axis up to 60 T under hydrostatic pressure up to 1.7 GPa. We find that both the lower (low-field) and the upper (high-field) boundaries of the first dome shift to higher fields with increasing pressure. In striking contrast, the summit of the dome is insensitive to pressure. Our study of the evolution of the Fermi surface pockets with pressure demonstrates that across the lower boundary, the BCS relation between the critical temperature and the density of states (set by the degeneracy of the Landau levels) remains valid under pressure. This weak-coupling behavior is disrupted at high magnetic field, when the critical temperature approaches a ceiling set by a parameter set by BEC, which shows little variation with pressure.

\section{Methods}

\begin{figure*}
	\centering
	\includegraphics[width=18cm]{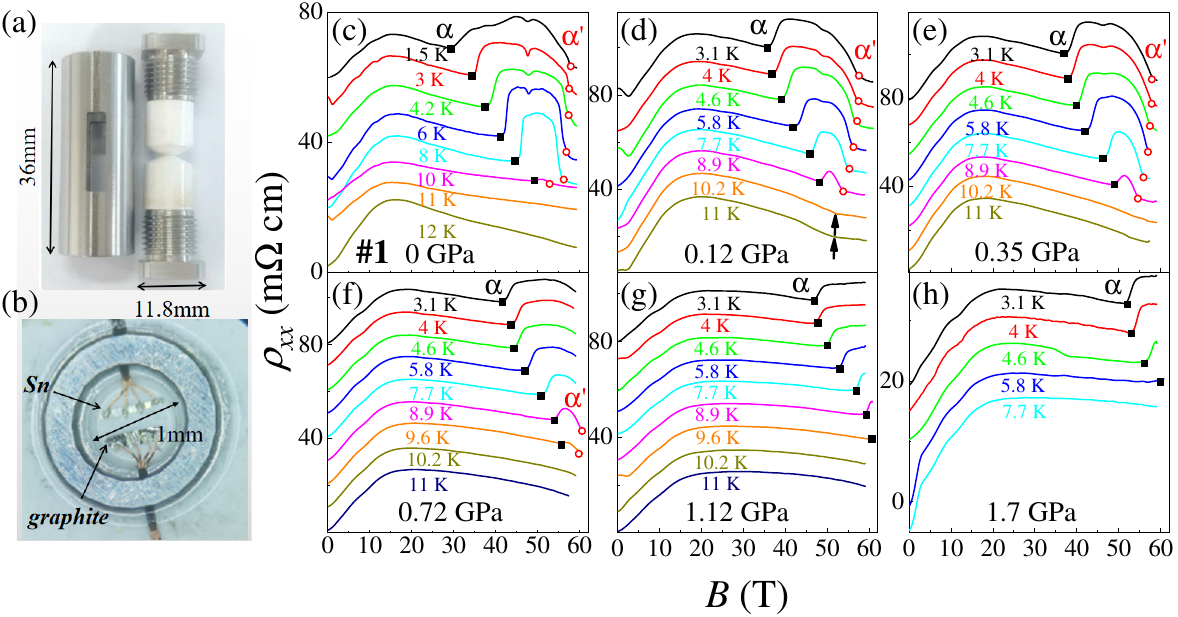}
	\caption{ \textbf{Pressure cell and magneto resistivity results :}
		(a) Photo of the pressure cell of external diameter 11.8 mm used in the pulsed-magnet (b) Photo of a Kish graphite sample in the pressure cell. The pressure was determined in-situ by the superconducting temperature transition of tin.
		(c)-(h) Field dependence of $\rho_{xx}$ up to 60 T at various temperatures and for different pressures. Curves are shifted for clarity. The onset transition ($\alpha$) and the re-entrant transition ($\alpha'$) of the first dome are indicated with black solid squares and red empty circles. The Shubnikov–de Haas oscillations at low-field are also observed at low temperatures (see supplemental material S$_2$). The $\alpha$ and $\alpha'$ are shifting to higher field after applied pressure. Under 1.12 Gpa, the $\alpha'$ shifts beyond 60 T.}
	\label{figure1-RB}
\end{figure*}

The pressure cell used in this study has been developed to fit in the pulsed field magnets of the Wuhan National High Magnetic Field Center. It is Bridgman type pressure cell adapted from the design of D. Braithwaite \textit{et al.} \cite{Braithwaite2016}. The cell body, with a diameter of 11.8 mm and a length of 36 mm, is crafted from MP35N (see Fig. \ref{figure1-RB}(a) for a photo of the cell). The anvils are machined of ZrO$_2$. Daphne 7373 was used as the pressure transmitting medium (see the supplement for additional details regarding the pressure cell). The sample space of diameter 1 mm can host a sample and a tin sample as shown on Fig. \ref{figure1-RB}(b). The superconducting transition temperature of the tin sample is used as an \textit{in-situ} measurement of the pressure in the cell. The magnetoresistance of graphite was measured with the standard four-probe method. The electrical current was applied in-plane and the magnetic field was applied along the $c$-axis for all samples. Sample temperature has been measured by a calibrated cernox thermometer attach to the body of the pressure cell. The unavoidable heating of the pressure cell during the pulse has been corrected through a comparison of the anomalies position with and without the gasket at ambient pressure (see supplement material section S$_1$).

%A calibrated cernox thermometer was attached to the body of the cell in order to measure the temperature of the sample. Note that the temperature of the samples  has been corrected by from the heating of cell body during the field pulse (see supplement material \textcolor{blue}{section S$_1$}). 

%The experiments under pulsed field were carried out down to 3 K and up to 60 T. All of the graphite samples used in this study are Kish graphite.

%?The cell was generated using a piston-clamp type pressure cell. 

\section{RESULTS}

% I think this sentence is not that useful : $\rho_{xx}$ shows a large magnetoresistance at low magnetic field due to the large mobility of the charge carriers in good agreement with \cite{Du2004}. 
%Under pressure, phase A shifts to higher magnetic field. 

Fig. \ref{figure1-RB}(c)-(h) shows the field dependence of the in-plane resistivity ($\rho_{xx}$) at various temperatures for pressures of 0, 0.12, 0.35, 0.72, 1.12 and 1.7 GPa (see the supplement material section S$_2$ for the Hall response). Curves are shifted for clarity. At zero pressure, see Fig. \ref{figure1-RB}(c), $\rho_{xx}$ displays a sudden increase above 20 T. This jump shifts to higher magnetic field as the temperature increases. Above 10 K, as reported previously \cite{Hiroshi2009}, the anomaly vanishes. The onset transition and the high field boundary, labelled $\alpha$  and $\alpha'$, following  previous authors \cite{Hiroshi2009}, are marked by black squares and red circles, respectively. The phase between $\alpha$ and $\alpha'$ is labelled the phase A (the first dome) \cite{LeBoeuf2017,Zhu2019}.

\begin{figure*}
	\includegraphics[width=18cm]{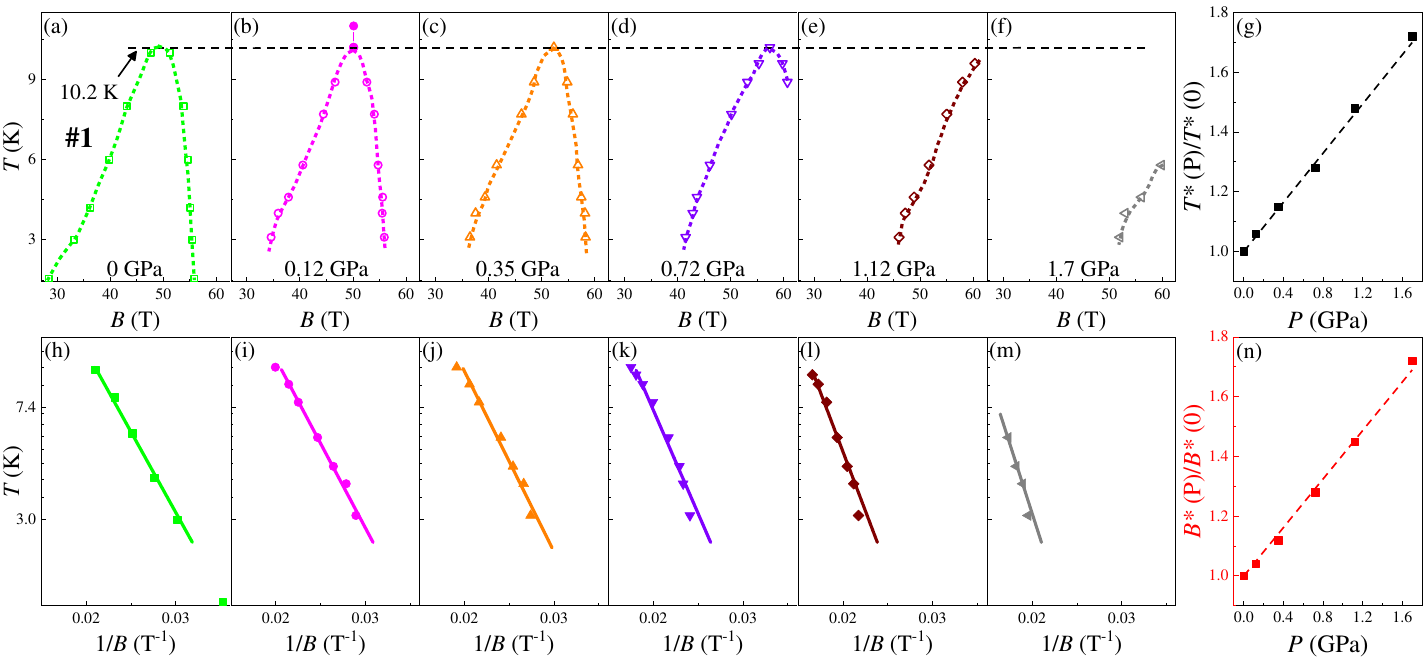}
	\caption{ \textbf{$T-B$ phase diagram of graphite under pressure :  }
		(a)-(f) $T–B$ phase diagrams for $H\parallel c$-axis at the pressure of 0, 0.12, 0.35, 0.72, 1.12 and 1.7 GPa. The dome is shifting to higher field under hydrostatic pressure. In contrast, the summit of the dome is independent of the pressure. The two solid circles in the (b) show the kinks in magnetoresistance.
		(h)-(m) $T~vs~1/B$ at different pressure. The solid lines are a fit of the low-field boundary of phase $\alpha$ using Eq.(\ref{EqBCS}), see the text.
		%	\st{(c) The $B–P$ phase diagrams for high-field boundary, the energy scale of depairing exciton characterized by the $\Delta B$ ($B$ (10.2 K)-$B$ (3.1 K)) keep almost unchanged with the pressure increases.
			%	(d) The critical temperature of BEC hardly changes with pressure.}
		(g)(n) Pressure dependence of the parameters $T^\ast$ and $B^\ast$ deduced from the fits.}
	\label{figure2-TB}
\end{figure*}

Under pressure, $\alpha$, $\alpha'$ , and therefore the phase A, shift towards higher magnetic fields. The evolution of the $T–B$ phase diagram with pressure is shown in Fig. \ref{figure2-TB}(a)-(f). Above 1.12 GPa, the high-field boundary $\alpha'$ moves above 60 T and exits our range of measurement. In contrast, the summit of the dome remains  at 10.2 K unchanged by the pressure, as indicated by the horizontal dashed line. %Pressure is thus a powerful knob to tune the boundaries of the field induced state of graphite.
%\textcolor{red}{WHAT IS THE INFORMATION GIVEN BY THIS SELF-EVIDENT BOASTFUL SENTENCE? IF YOU WANT TO KEEP IT, DELETE AT LEAST 'POWERFUL'. REPLY: I have deleted this sentence. }

%This behavior is almost the same at the measured pressure, but the phase A is shifting to higher fields after applied pressure. The high-boundary $\alpha'$ moves out of our measured field range (60 T here) at 1.12 GPa. The high-boundary $\alpha'$ is also closely out of our measured field range at 1.7 GPa. 

Black arrows on Fig. \ref{figure1-RB}(d) and solid circles in Fig. \ref{figure2-TB}(b) indicate the kinks in $\rho_{xx}$ which survives above 10 K. A similar anomaly at ambient pressure, above the field induced state, was detected in measurements of the sound velocity  \cite{LeBoeuf2017}, the out-of-plane magnetoresistance ($\rho_{zz}$) \cite{Zhu2017} and  the Nernst effect \cite{Wang2020}. This kink marks the field at which electron and hole Landau sub-bands simultaneously cross the Fermi level \cite{Zhu2017,Wang2020}, creating the most favorable conditions for an electronic instability such as an exciton condensation. 

%The solid circles in Fig. \ref{figure2-TB}(b) indicate the position of the kinks, which were marked by the black arrows in Fig. \ref{figure1-RB}(d).

%similar to theoretical calculations for EI \cite{Jerome1967,kozlov1965metal}. 
%\textcolor{blue}{
	%The solid circles in Fig. \ref{figure2-TB}(b) indicate the position of the kinks, which were marked by the black arrows in Fig. \ref{figure1-RB}(d).}

%The $\alpha$ and $\alpha'$ are shifting to higher field under pressure. 

\section{DISCUSSION}
\subsection{BCS regime under pressure}

At ambient pressure, in the low-field boundary of the phase A, the critical temperature ($T$) and field ($B$) follows:
\begin{equation}
	T(B)=T^\ast{\rm{exp}}(-B^\ast/B)
	\label{EqBCS}
\end{equation}

where $T^\ast$ and $B^\ast$ are adjustable parameters. This formula mimics a BCS-type expression: $k_BT_c(B)=1.14E_F {\rm{exp}}(-\frac{1}{N(E_F)V}) $, where $N(E_F)$ is the density of states (DOS) at the Fermi energy ($E_F$) and $V$ is the pairing interaction \cite{Iye1990,Yaguchi1993}. In this framework, $T^\ast$ is proportional to the Fermi energy, $B^\ast$ is inversely proportional to the $N(E_F)V$ product. The change of the critical temperature with the magnetic field is due to field dependence of the DOS which increases linearly with the magnetic field, driven by the degeneracy of the Landau levels \cite{Marcenat2021}. With increasing magnetic field, both the DOS and the critical temperature increase and approach the summit of the dome. 

Fig. \ref{figure2-TB}(h)-(m) show the  evolution of $T$ $vs$ $B^{-1}$ with pressure. For all studied pressures, Eq. \ref{EqBCS} is satisfied which allows a determination of $T^\ast$ and $B^\ast$. Fig. \ref{figure2-TB}(g) and (n) show the pressure dependence of $T^\ast$ and $B^\ast$ normalized by the ambient pressure values. Both quantities increase linearly with the pressure. Their slope is similar: $a=0.41\pm0.02$ GPa$^{-1}$ for $T^\ast(P)$ and $a=0.4\pm0.03$ GPa$^{-1}$ for $B^\ast(P)$.

To quantify the change of the Fermi surface induced by the pressure we studied the evolution of the Shubnikov-de Haas (SdH) oscillations in DC field up to 16 T and 1.7 GPa (see supplementary material section S$_3$). Fig. \ref{figure4-n+m}(a) shows the evolution of the SdH frequencies ($F$) and the effective mass $m^\ast$  deduced from their temperature dependence. The normalised pressure dependence of the Fermi energy of electrons ($E_{F,e,\perp}$), holes ($E_{F,h,\perp}$) and their average ($E_{{F,ave}}=(E_{F,h}+E_{{F,e}})/2$) are shown in Fig. \ref{figure4-n+m}(c). The $\partial {\rm{ln}}m^\ast/\partial P$ and $E_{{F,ave}}$ increase linearly with pressure with an slope of $a=0.38\pm0.03$ GPa$^{-1}$, in good agreement with an early and comprehensive quantum oscillation analysis by Brandt ($a=0.43\pm0.03$ GPa$^{-1}$)\cite{Brandt1980}, see the supplementary material section S$_4$. 

%only 3\% different from the results calculated from the SWM model by considering $\gamma_2$ using $E_F(P)=(1+aP)E_F(0)$\cite{Iye1990} with $a=0.38\pm0.03$ GPa$^{-1}$. This is consistent with the $a$ previously obtained by fitting $B^*$ and $T^*$ and with a previous report by Brandt ($a=0.43\pm0.03$ GPa$^{-1}$)\cite{Brandt1980}.

\begin{table}
	\centering
	\begin{tabular}{c|c|c}
		\hline
		&Methods & $a$ (GPa$^{-1}$)   \\ \hline
		
		Brandt \textit{et al.}& $\partial {\rm{ln}}m^\ast/\partial P$ & $0.43\pm0.03$ \cite{Brandt1980} \\ 
		(1.7 GPa, 2 K)& &\\
		\hline
		
		Iye \textit{et al.}& $T^\ast(P)=(1+aP)T^\ast(0)$ & $0.29\pm0.01$ \cite{Iye1990} \\ 
		(1.05 GPa,$<$ 1.5 K)&$B^\ast(P)=(1+aP)B^\ast(0)$&$0.29\pm0.01$ \cite{Iye1990}\\
		\hline
		&$\partial {\rm{ln}}m^\ast/\partial P$  &  $0.38\pm0.02$     \\

		Present work&$T^\ast(P)=(1+aP)T^\ast(0)$ &  $0.41\pm0.03$   \\
		
		(1.7 Gpa, 3 K-10 K) &$B^\ast(P)=(1+aP)B^\ast(0)$ & $0.4\pm0.02$   \\
		
		&$E_F(P)=(1+aP)E_F(0)$ & $0.38\pm0.03$   \\
		
		\hline
		
	\end{tabular}
	\caption{Coefficient of the linear pressure dependence of $\partial {\rm{ln}}m^\ast/\partial P$, $T^\ast$, $B^\ast$ and $E_F$ according to \cite{Iye1982,Brandt1980} and our study.}
	\label{methods}
\end{table}

%Comparison of the logarithmic derivatives, the $a$ (in GPa$^{-1}$) was around 0.4 Gpa$^{-1}$ obtained by four various methods. The $a$ deduced from BCS for low-field boundary of first is consistent with these from the quantum oscillations method below quantum limit which is also as reported by Brandt \textit{et al.} \cite{Brandt1980}, pointing to the validation of BCS picture for the low-field boundary.

%Comparison of the logarithmic derivatives, the $a$ (in GPa$^{-1}$) was around 0.4 Gpa$^{-1}$ obtained by four various methods. The $a$ deduced from BCS for low-field boundary of first is consistent with these from the quantum oscillations method below quantum limit which is also as reported by Brandt \textit{et al.} \cite{Brandt1980}, pointing to the validation of BCS picture for the low-field boundary.
%Now let's link this striking common pressure dependence to the band structure of graphite described by 
%the Slonczewski-Weiss-McCure (SWM) tight binding model. 

Table \ref{methods} summarizes the amplitude of the pressure dependence of the four quantities studied : $T^*(P)$, $B^*(P)$, $E_F(P)$ and $\partial {\rm{\ln}}(m^\ast)/\partial P$ (see the supplemental information S$_4$). Remarkably they display the same pressure dependence. This striking observation can be linked to the pressure dependence of a single parameter ($\gamma_2$) of the Slonczewski-Weiss-McCure (SWM) tight-binding model of the band structure of graphite. This model is formed by seven energy scales ($\gamma_i;i=0-5$ and $\Delta$) \cite{McClure1957} that represent interactions between neighboring carbon atoms. The parameter $\gamma_2$ quantify the inter-layer coupling between the two sub-lattices. It sets the $c$-axis dispersion : $E(k_z)$=-2$\gamma_{2}\sin(\frac{c_{0}k_{z}}{2})$ where $c_{0}=2c$ with $c$ is the interlayer distance and $k_z$ is the $c$-axis momentum \cite{Iye1982}. Under pressure the inter-layer coupling and $\gamma_2$ increase linearly with pressure : $\gamma_2(P)=(1+aP)\gamma_2(0)$ with $a$ ranging from 0.23 to 0.43GPa$^{-1}$ according to various experiments done at different temperatures \cite{Brandt1980,Iye1990}. Thus, the pressure impacts $T^\ast$, which is approximately proportional to $E_F$, and  $B^\ast$, which scales inversely with $N(E_{F})V$ if the pairing interaction $V$ does not change significantly. Our findings suggest that the pressure-induced variation in $\gamma_2$ is not only the driving force behind the linear increase in $E_F$,  but also in $N(E_{F})$ (as detailed in supplemental material S5) \cite{Iye1990}.

\subsection{BCS-BEC cross-over}

\begin{figure}
\includegraphics[width=8.5cm]{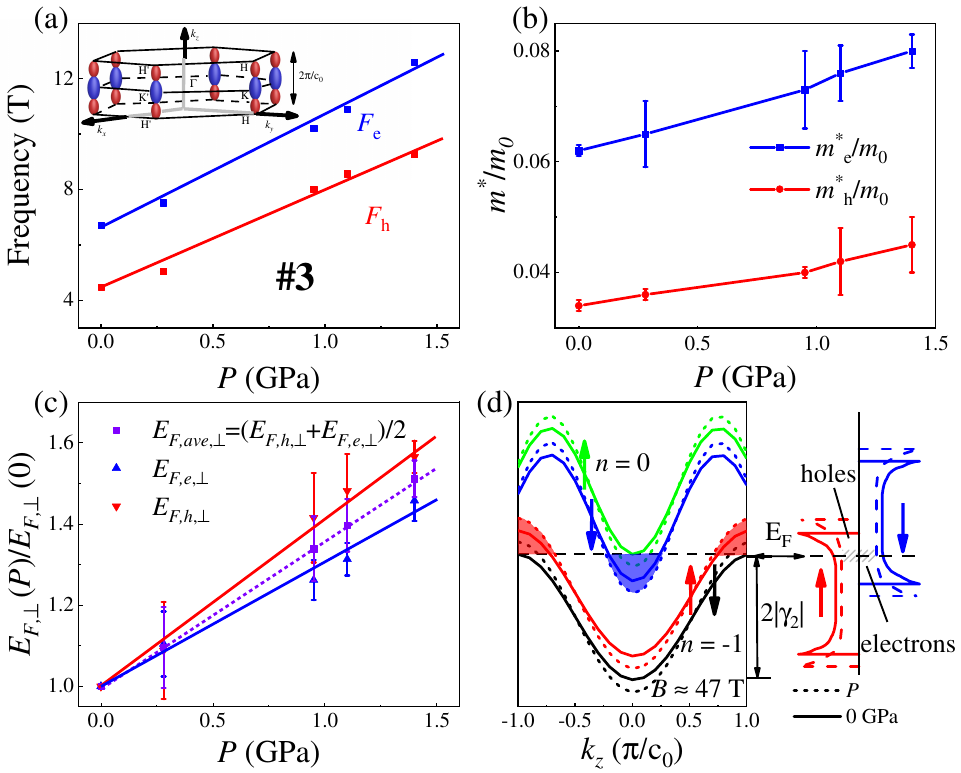}
\caption{ \textbf{Pressure dependence of the Fermi surface properties of graphite:}
(a) Pressure dependence of the SdH frequencies ($F$) of the electrons (blue) and holes (red). Insert :  sketch of the Fermi surface of graphite, formed by six adjacent ellipsoid pockets (electron in blue and hole in red). (b) Pressure dependence of the effective masses $m$ of the electrons (blue) and holes (red). (c) Pressure dependence of the Fermi energy of the electrons, holes and its average (purple points).
(d) Sketch of the Landau-level spectrum of graphite close to the summing of the dome. The dashed-solid curves represent the conditions at ambient pressure, while the dashed curves correspond to those under pressure. On the right, the corresponding density of states is depicted.}
\label{figure4-n+m}
\end{figure}

\begin{figure}
\includegraphics[width=8cm]{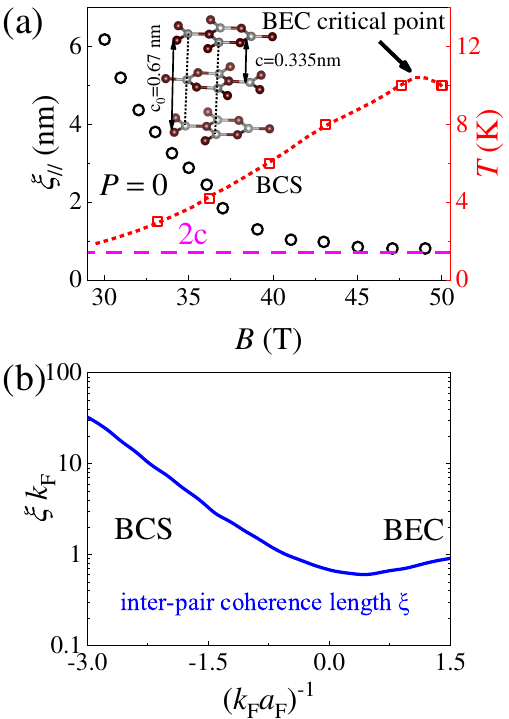}
\caption{ \textbf{BCS-BEC crossover in graphite:} 
(a) Field dependence of $\xi_\parallel$ (black open circle points, see the text for the definition), at ambiant pressure compare with the field dependence of the critical temperature ($T$) in red square points. The purple dashed line is the A-A interlayer distance $c_0$. When $\xi_\parallel$ saturates to $c_0$, $T$ saturates also at its largest value. The inset shows the lattice structure of graphite. (b) The inter-pair coherence length $\xi k_F$ (blue full line) at the mean-field level {\it{vs.}} the coupling parameter $(k_Fa_F)^{-1}$. Reproduced from Ref.\cite{Strinati2018}. 
}
\label{figure6-xi}
\end{figure}

%Fig. \ref{figure3-EF}(d) summarizes the $T–B$ phase diagrammatic sketch with and without hydrostatic pressure. The first dome is shifting to higher field under pressure and the critical temperature keep almost unchanged. The low-field boundary evolving to a mean-field expression for critical temperature as show in the Fig. \ref{figure2-TB}(h)-(m)\cite{Hiroshi2009}, is resolved. On the high-field side, the exciton binding energy becomes lower than the band gap, leading to the exciton depairing\cite{Wang2020}. The magnetic field energy where the peak position to the exciton despairing position is roughly constant.  

%Under the BCS-BEC crossover scenario, we can easily understand this stable critical temperature with pressure.

%In the BCD picture $T_c$ increases  

In contrast to the low-field boundary regime that is tuned by the pressure, the maximum critical temperature of the dome is independent of it. This result points to two distinct regimes in the dome. It was recently noticed that the summit of the dome at ambient pressure, which occurs at $\simeq$ 10 K, is close to the degeneracy temperature of excitons \cite{Wang2020}. Indeed the inter-plane distance between excitons and  the interplane thermal de Broglie wavelength match each other \cite{Wang2020} at this temperature, indicating that this summit corresponds to the BEC temperature. The present results imply that this critical temperature does not show any detectable shift with pressure despite the pressure-induced change of the Fermi temperature.  

%Fig. \ref{figure3-EF}(d) summarizes the $T–B$ phase diagrammatic sketch with and without hydrostatic pressure. The first dome is shifting to higher field under pressure and the critical temperature keep almost unchanged. The low-field boundary evolving to a mean-field expression for critical temperature as show in the Fig. \ref{figure2-TB}(h)-(m)\cite{Hiroshi2009}, is resolved. On the high-field side, the exciton binding energy becomes lower than the band gap, leading to the exciton depairing\cite{Wang2020}. The magnetic field energy where the peak position to the exciton despairing position is roughly constant.  Under the BCS-BEC crossover scenario, we can easily understand this stable critical temperature with pressure.

In order to understand why the summit of the dome is independent of the pressure, let us now put it in the context of the crossover between the BCS to BEC regime. In the weak coupling  (BCS) limit,  the coherence length ($\xi$) is much longer than the distance between e-h pairs ($d$), allowing the applicability of a mean-field BCS-type formula linking the critical temperature to the density of states. Graphite being an anisotropic material, both length scales are anisotropic:  $d_\perp \gg d_{\parallel}$ (the interparticle distances in the basal plane $d_\perp=19.5$ nm \cite{Wang2020} and $d_\parallel \approx 1$ nm \cite{Marcenat2021}) and $\xi_\perp \ll \xi_{\parallel}$. In contrast to $d_\perp$, which depends on carrier concentration, $d_{\parallel}$ is set by the distance between layers. One can also estimate $\xi_\parallel$, the coherence length along the $c$-axis and compare it with $d_{\parallel}$ in order to see how the system evolves from the weak limit ($\xi_\parallel > d_{\parallel}$) to the strong limit ($\xi_\parallel \simeq d_{\parallel}$).

Fig. \ref{figure6-xi}(a) shows the evolution of $\xi_\parallel$ with magnetic field using the BCS formula $\xi_\parallel=\frac{\hbar^2k_{F,\parallel}}{\pi m_\parallel^\ast \Delta_c}=\frac{\hbar v_{F,\parallel}}{\pi \Delta_c}$. Here, $\Delta_c$ is the energy gap measured by out-of-plane resistance measurements \cite{Fauque2016}. It increases with the magnetic field. Assuming $k_{F,\parallel}$ to be $\approx\frac{\pi}{8c}$ \cite{Arnold14,Marcenat2021}, allows one to extract  $\xi_\parallel$ and see that its steady decrease  with increasing magnetic field decelerates first and then saturates to 2$c$ (in other words, $d_{\parallel}$). Thus, there is an upper bound to the critical temperature, because the coherence length cannot become shorter than the interbosonic distance. This picture is to be compared with the theoretical picture of the BCS–BEC crossover shown in Fig. \ref{figure6-xi}(b). In the weak-coupling BCS regime \cite{Strinati2018}, the inter-pair coherence length $\xi$ decreases. In the strong-coupling BEC regime, when $(k_Fa_F)^{-1} \geq 1$\cite{Strinati2018}, the interaction increases further, but $\xi$ ceases to decrease. This is consistent with our observation of the decrease in $\xi_{\parallel}$ followed by its saturation. Furthermore, at the BCS-BEC crossover, $\xi k_F$ shows a minimum at $\approx 0.6$. In the case of graphite, this corresponds to $\xi=1.53c$, broadly consistent with the saturation of $\xi$ at $\approx 2c$ found in Fig. \ref{figure6-xi}(a).

%This analyse is quantitatively support by the position of the minimum of $\xi k_F$ $\approx 0.6$ at the BCS-BEC crossover. 

%hich is consistent with the $c$-axis distance $2c$.

How does this picture evolve with pressure?  The short answer to the question is that the pressure leaves $2c$ almost unchanged (it changes by less than 3 \% at 1 GPa \cite{Brandt1988}). Since the out-of-plane correlation length cannot become shorter than 2$c$, the bound to the BCS critical temperature remains identical despite the shift in the parameters. For a more comprehensive answer, one needs to quantify $\xi_\parallel$ under pressure. This requires measuring $\Delta_c$. Let us note however, that the change in $v_{F,\parallel}(P)$, inferred from fermiology, is small. It decreases by less than $\approx$ 10\%, as a consequence of the decrease in both the effective mass and the Fermi radius (see the supplement material S$_3$).

In the BCS-BEC crossover, the hierarchy between normalised chemical potential and order parameter change, without altering the ground state and causing any phase transition \cite{Parishbook}. The crossover is achieved by changing  the  ratio of the size of the pairs and the distance between the particles. Therefore, it has been argued \cite{Parishbook} that to drive the crossover, one can either change the particle density or the amplitude of the fermion-fermion interaction. The latter road (`interaction driven') is taken in the atomic gases with Feshbach resonance \cite{Jochim2003BCSBEC,Greiner2003BCSBEC,Zwierlein2003BCSBEC,Regal2004BCSBEC,Kinast2004BCSBEC,Bourdel2004BCSBEC,Zwierlein2004BCSBEC,Bartenstein2004BCSBEC,Parishbook}. The former road  `density driven'  was theoretically invoked for excitons decades ago \cite{Comte1982}, but is hard to realize experimentally. Graphite under a strong magnetic field offers an alternative. Our result shows that pairing interaction $V$ is almost pressure independent. It is the increase of DOS, induced by the magnetic field, that drives here the BCS-BEC crossover and not the tuning of the pairing interaction or the particle density.

%Graphite under a strong magnetic field offers an alternative. \textcolor{blue}{In graphtie, the DOS which is proportional to the magnetic field and is therefore the driver for the crossover. Since the pairing interaction $V$ is almost pressure independent and the change in carrier concentration is small.} This is \textcolor{blue}{also a case in which this simple dichotomy vanishes because of a large underlying anisotropy. Both relevant length scales are strongly anisotropic. The magnetic field slightly alters the particle density, causes a shift in the chemical potential, but most significantly, it augments the DOS.}
%amplifies interaction by weakening the screening, but also
%\textcolor{red}{Lastly, it is noteworthy that the competition among the lowest Landau levels may introduce additional complexity in the occurrence of high magnetic transitions.} 

Lastly, it is noteworthy that valley \cite{Yoshioka1981} and orbital \cite{Ho2011} degree of freedom can introduce additional complexity in the high magnetic field regime of graphite. Recently, a theoretical study  by Kousa, Wei and Macdonald  found that  the $n=0$ and $n=1$  Landau levels of bilayer graphene are sensitive to the details of the particle-hole symmetry breaking and concluded that the mixing of Landau orbitals may affect the physics of bulk graphite at high magnetic fields \cite{kousa2024orbital}. The link between two research fields, field-induced electron-hole pairing  in 3D graphite and fractional quantum Hall effect in 2D graphene remains a totally unexplored territory. 

\textcolor{red}{
}
%In other words, from $\xi_\parallel \to d_\parallel$ to $\xi_\parallel \to c$ maybe need more large magnetic field, as show in the Table. \ref{tab:my_label}.  

%\begin{table}
%\    \centering
%\\begin{tabular}{c|c|c}
%\\hline
%\ & 0 GPa & P ( $\approx$ 0.7 GPa)\\ \hline
%\$\xi_\parallel \gg d_\parallel$ & 33 T & 41 T \\
%\$\xi_\parallel \to d_\parallel$ & 47 T & 58 T\\
%\$\xi_\parallel \to c$ &  \\
%\\hline
%\\end{tabular}
%\    \caption{When $\xi_\parallel \gg d_\parallel$, it start to leave the weak-coupling BCS regime, the BCS-BEC crossover occurs when $\xi_\parallel \to d_\parallel$, at last, when $\xi_\parallel \to c$ the field reaches the critical point.}
%\    \label{tab:my_label}
%\\end{table}

\section{CONCLUSIONS}

In summary, we performed a study of magnetoresistance of Kish graphite up to 60 T under pressure up to 1.7 GPa. The $\alpha$ and $\alpha'$ transitions shift to higher fields, while the summit of the dome remains at the same temperature. We argued that this observation can be understood by considering the BCS parameters of the low field transitions and the BCS-BES crossover constraints at the summit of the dome.

\section{ACKNOWLEDGMENTS}
This work was supported by The National Key Research and Development Program of China (Grant No.2022YFA1403500), the National Science Foundation of China (Grant No.12004123, 51861135104 and No.11574097) and  the Fundamental Research Funds for the Central Universities (Grant no. 2019kfyXMBZ071). K. B was supported by the Agence Nationale de la Recherche (ANR-19-CE30-0014-04). B. F. was supported by the Agence Nationale de la Recherche (ANR-18-CE92-0020-01) and by Jeunes Equipes de l$'$Institut de Physique du Coll\`ege de France. X. L. acknowledges the China National Postdoctoral Program for Innovative Talents (Grant No.BX20200143) and the China Postdoctoral Science Foundation (Grant No.2020M682386).

\noindent
* \verb|zengwei.zhu@hust.edu.cn|\\


%The Fermi surface increase with pressure, which makes the quantum limit shift to higher field.  With the pressure is increased, the increase in the magnetic field needed to enter and destroy the exciton phase, which the electron spin-up and the hole spin-down subbands simultaneously cross the Fermi level with the necessary magnetic field is increased. So we observed that the $\alpha$, $\alpha'$ and dome are shifting to higher field with pressure. 

 %Our previous study discovered that the behavior of the peak at 47T is consistent with the characteristics of quantum oscillation by Nernst measurements in pulsed-magnetic-field as same as the quantum limit (7.4T). 
\clearpage

%\title{The field-induced insulating dome of graphite under pressure }

% Add 'S' to the numbering inside the appendix
\renewcommand{\thesection}{S\arabic{section}}
\renewcommand{\thetable}{S\arabic{table}}
\renewcommand{\thefigure}{S\arabic{figure}}
\renewcommand{\theequation}{S\arabic{equation}}

\setcounter{section}{0}
\setcounter{figure}{0}
\setcounter{table}{0}
\setcounter{equation}{0}

{\large\bf Supplemental Materials for "Tuning the BCS-BEC crossover of electron-hole pairing with pressure"}

\section{Temperature calibration and data reproducibility}

As described in Method, the pressure cell is made of MP35N which has a low thermal conductivity. Yet, during the magnetic field pulse the cell heats due to the unavoidable induced eddy currents. To minimise their effects we show in this manuscript the data collected during the rising field. 

%Furthermore to quantify the change in temperature induced by the pressure cell we use the position in field of the $\alpha$ transition, that is an extremely sensitive in-situ thermo-meter to the temperature ($\frac{dTc}{dB} \approx$). 

\begin{figure}[htp]
	\includegraphics[width=8.5cm]{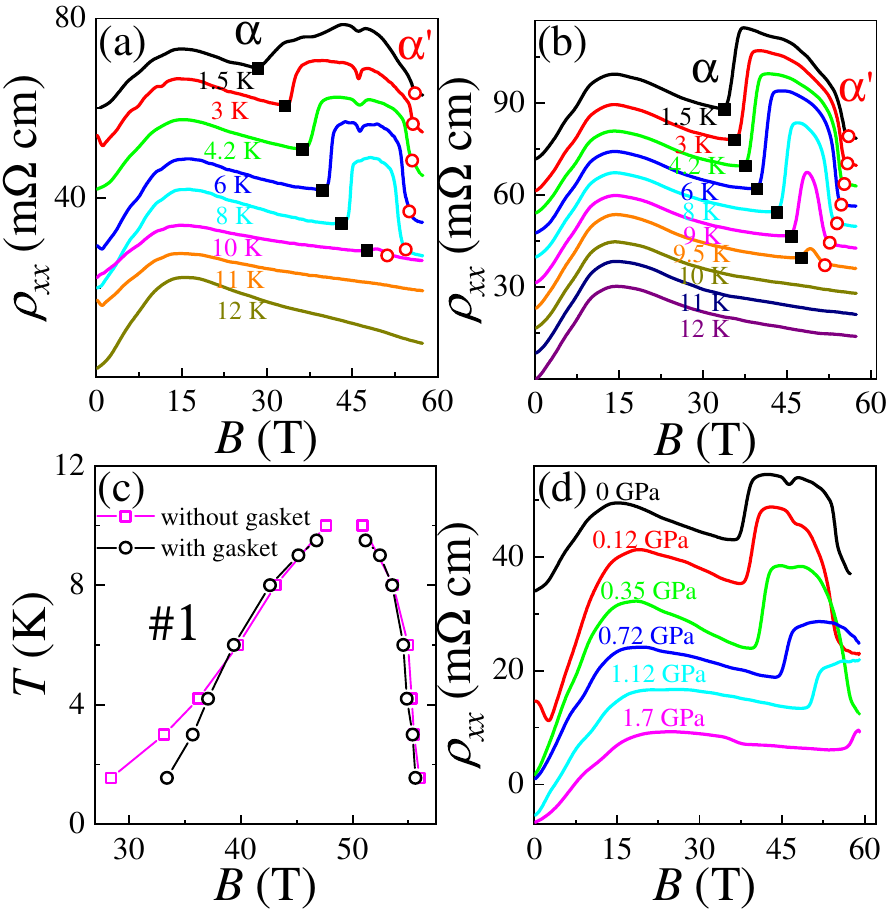}
	\caption{ %\textbf{Crystal structure, resistivity }
		\textbf{Temperature calibration} (a) and (b) The field dependence of the in-plane magneto-resistance ($\rho_{xx}$) at different temperatures for $B \parallel$ $c$-axis with gasket or without gasket at 0 GPa in the same sample \#1. Data are shifted for clarity. (c) The $T–B$ phase diagrams for $B \parallel$ $c$-axis with gasket or without gasket at 0 GPa. (d) The magnetic field dependence of resistivity at 4.2 K with different pressure.}
	\label{s-figure2}
\end{figure}

To calibrate the change of the temperature induced by the pressure cell on the samples, we utilize the critical field of $\alpha$ in graphite, that is an extremely sensitive \textit{in-situ} thermometers ($\frac{dT_c}{dB}=\frac{T_cB^\ast}{B^2}\approx$ 0.2 K/T @($T=1.5$ K)). We conducted measurements of $\rho_{xx}$  in the pressure cell, at ambient pressure and up to 60 T, both with and without the gasket in the same sample \#1, see Fig. \ref{s-figure2}(a) and (b). Notably, the gasket is the primary source of heating.

%Although the designed window effectively reduces the metal area and the temperature rise near the sample, we The main heating source comes from the gasket by stainless steel. Measuring $\rho_{xx}$ in the pressure cell at 0 GPa up to 60 T with gasket and without gasket to qualify the heating effect under pulsed field. The temperature in the measurement with gasket under pressure is corrected to the temperature in the measurement without gasket.

%Although the designed window significantly reduces the metal area and temperature increase near the sample, the primary heating source remains the stainless steel gasket. To accurately assess the heating effect under pulsed field conditions, 

%Fig. \ref{s-figure2}(a) and (b) show the field dependence of the in-plane magnetoresistance ($\rho_{xx}$) at different temperatures for $H\parallel$ c-axis with gasket or without gasket at 0GPa. It is shows clearly the relationship between the critical temperature and the critical magnetic field at the left boundary of phase $\alpha$ and the right boundary of phase $\alpha'$. The B-T phase diagram for $H\parallel$ c-axis in Fig. \ref{s-figure2}(c). The phase $\alpha$ is shifting to higher field below 6K with gasket, The main heating source comes from the gasket. Thus, the temperature in the measurement with gasket under pressure should be to be corrected to the temperature in the measurement without gasket. The results are shown in the Table \ref{temperature}.

%Subsequently, the temperature measurements obtained with the gasket under pressure were adjusted to align with those taken without the gasket, ensuring accurate temperature correction.

%Notably, the gasket is the primary source of heating. Therefore, it is essential to adjust the temperature measurements obtained with the gasket under pressure to align with those taken without the gasket.

%Fig. \ref{s-figure2}(a) and (b) shows $\rho_{xx}$ as a function of field at various temperatures for $B\parallel$ $c$-axis, comparing measurements with and without the gasket at 0 GPa. 
%C
%These figures clearly demonstrate the correlation between the critical temperature and the critical magnetic field at the left boundary of phase $\alpha$ and the right boundary of phase $\alpha'$. 

\begin{figure*}[htp]
	\includegraphics[width=18cm]{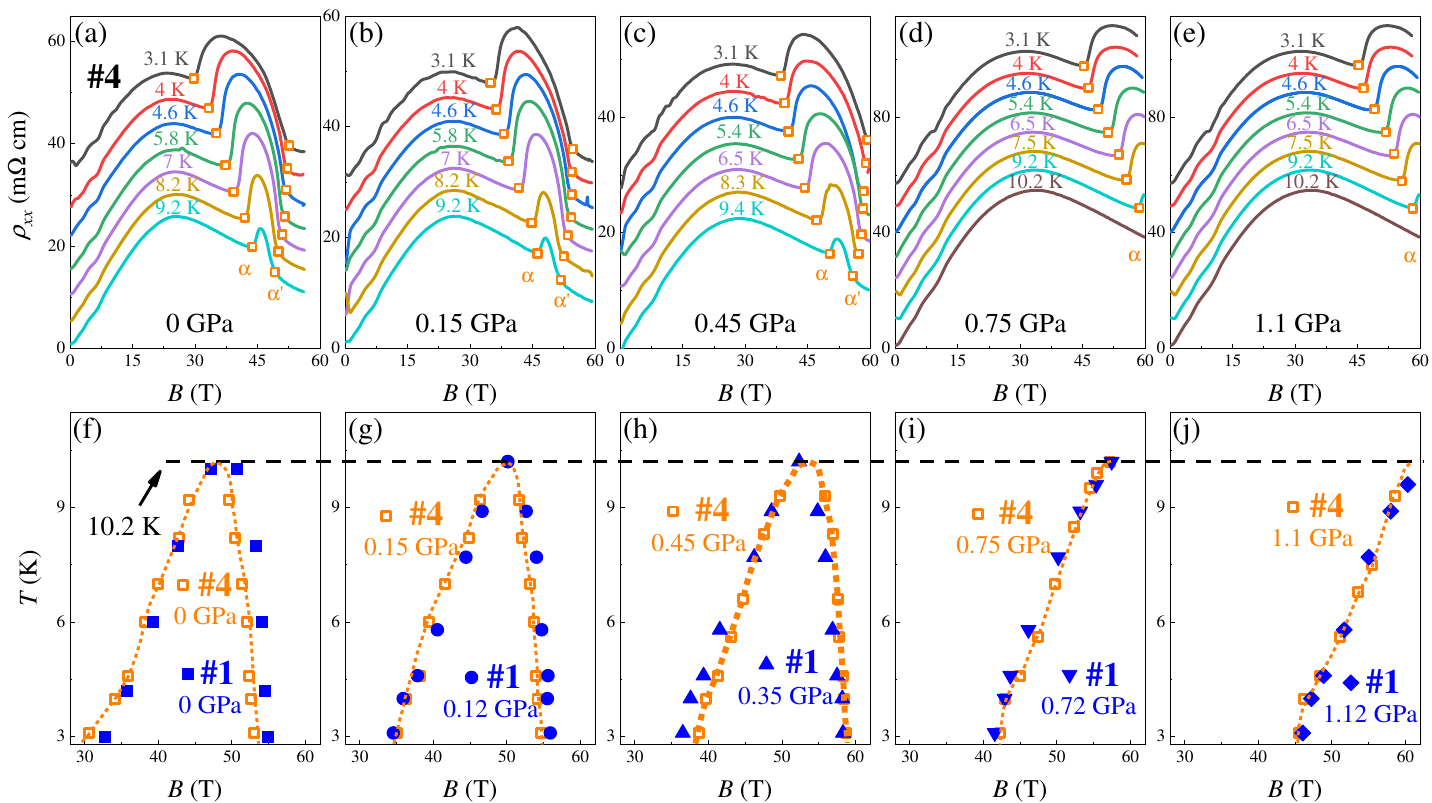}
	\caption{ %\textbf{Crystal structure, resistivity}
		\textbf{\textit{T-B} phase diagram for the sample \#4} 
		(a)-(e) The field dependence of $\rho_{xx}$ at various temperatures up to 60 T for different pressures for sample $\#4$. Curves are shifted for clarity. 
		(f)-(j) The $T–B$ phase diagrams for sample $\#4$ (open orange symbols) for five pressures compared with sample $\#1$ (closed blue symbols). 
	}
	\label{s-figure7}
\end{figure*}

The deduced \textit{T-B} phase diagram with and without the gasket is shown in Fig. \ref{s-figure2}(c). Below 6 K the measurements with the gasket is clearly at a higher temperature than without. We therefore adjust the temperature obtained with the gasket under pressure to align with those taken without the gasket. The corrected temperature values are summarized in Table \ref{temperature}. The data shown in the main text have been collected on sample \#1. Measurements of $\rho_{xx}$ in an other sample, labelled \#4, are shown on Fig.\ref{s-figure7}(a)-(e). The pressure of the \textit{T-B} phase diagram is in excellent agreement with sample \#1, see Fig.\ref{s-figure7}(f)-(j), which demonstrate the reproducibility of the results.

\begin{figure*}[htp]
	\includegraphics[width=18cm]{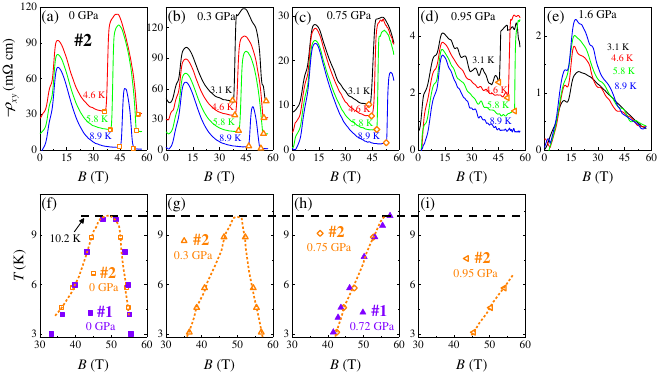}
	\caption{ %\textbf{Crystal structure, resistivity }
		\textbf{The Hall effect under pulsed-magnetic-field}
		(a)-(e) The in-plane longitudinal Hall resistance at different temperature comprised between 3.1 K to 8.9 K up to 60 T under pressure up to 1.6 GPa. 
		(f)-(i) The $T–B$ phase diagrams for $H\parallel c$-axis under different pressure from the Hall resistivity of \#2 ( open oragne symbols), and compared with the sample $\#1$ (closed purple symbols).
	}
	\label{s-figure3}
\end{figure*}

\begin{table}
	\centering
	\begin{tabular}{c|c|c|c|c|c|c|c|c}
		\hline
		with gasket (K) & 1.5 & 3 & 4.2 & 6 & 8 & 9 & 9.5 & 10 \\ \hline
		without gasket (K) & 3.1 & 4 & 4.6 & 5.8 & 7.7 & 8.9 & 9.6 & 10.2 \\ \hline
	\end{tabular}
	\caption{The different temperatures with gasket or without gasket. The temperature in the measurement with gasket under pressure should be to be corrected to the temperature in the measurement without gasket.}
	\label{temperature}
\end{table}

\section{Hall measurement under pulsed-magnetic-field}

In addition to $\rho_{xx}$, Hall resistivity ($\rho_{xy}$) have been measured up to 60 T at various temperatures and for pressure $P$ = 0, 0.3, 0.75, 0.95 and 1.6 GPa in the sample \#2. In order to extract $\rho_{xy}$ measurements of the transverse voltages have been done for positive and negative magnetic field followed by an anti-symetrisation. 

%two samples, labelled \#2 and \#4 respectively

Fig. \ref{s-figure3}(a)-(e) shows the field dependence of $\rho_{xy}$ under these pressure. The overall amplitude of $\rho_{xy}$ decreases as the pressure increase due to the increase in the carrier density confirmed by quantum oscillations measurements, see next section. Like in $\rho_{xx}$, the entrance (and re-entrance) of the field induced state is marked by a sharp increase (decrease) in $\rho_{xy}$ that shift to higher magnetic field once the temperature or the pressure increase. Fig. \ref{s-figure3}(f)-(i) show the \textit{T-B} phase diagrams deduced from $\rho_{xy}$ in sample \#2 compare with to the one deduced from $\rho_{xx}$ in sample \#1. Both type of measurements provide the same results. We note that the summit of the dome remains unchanged and consistent among all the samples studied.

\section{The Shubnikov-de Haas effect and Hall effect analyse in low-field}

\begin{figure*}[htp]
	\includegraphics[width=17cm]{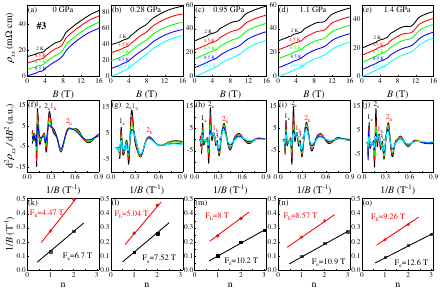}
	\caption{ %\textbf{Crystal structure, resistivity}
		\textbf{The Shubnikov de Haas(SdH) oscillations} (a)-(e) The magnetoresistance as a function of the magnetic field at different temperatures for $B\parallel$ $c$-axis under different pressure. 
		(f)-(j) The oscillatory part of resistivity $d^2\rho_{xx}/dB^2$ for $H\parallel$ c-axis. 
		(k)-(o)   The Landau fan diagram derived from the field position of the distinct peaks of  $d^2\rho_{xx}/dB^2$ for both the hole and electron pockets.}
	\label{s-figure4}
\end{figure*}

\begin{figure*}[htp]
	\includegraphics[width=17cm]{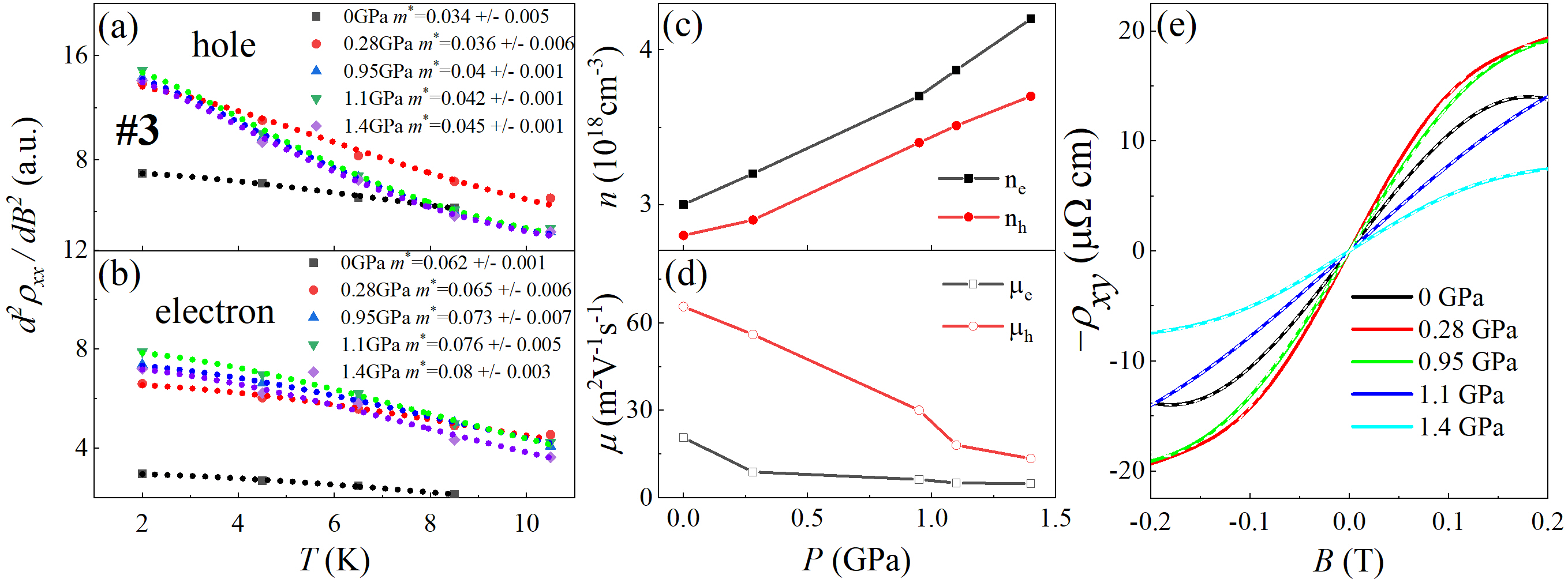}
	\caption{ %\textbf{Crystal structure, resistivity}
		\textbf{Pressure dependence of the the effective masses, carrier density and mobility : }
		(a)(b) The L-K fitting with the effective masses of electron and hole at different pressure. 
		(c)(d) The pressure dependence of carrier density and mobility with the pressure increases.
		(e) Hall resistivity under different pressure in low temperature. The dotted line represents a two-band fit to the data using the formula \ref{Rxy}.}
	\label{s-figure5}
\end{figure*}

\paragraph{Analyse of SdH oscillations}

In order to quantify the change of the Fermi surface as function of the pressure ($P$), we measured the field dependence of $\rho_{xx}$ and $\rho_{xy}$ of graphite in DC fields with PPMS (Physical Propertie Measurement Sysytem) and Oxford Instrument Integra 16 T system at $P$ = 0, 0.28, 0.95, 1.1 and 1.4 GPa, see Fig. \ref{s-figure4}(a)-(e). On top of a monotonous large magneto-resistance quantum oscillations in $\rho_{xx}$, known as the Shubnikov-de Haas (SdH) effect, are observed. Study of these oscillations allow to quantify the change of the Fermi surface (effective masses and frequencies) with the pressure. The trace of these oscillations are clearly visible in the second derivative of $\rho_{xx}$, see Fig. \ref{s-figure4}(f)-(j). Fig.\ref{s-figure4}(k)-(o) show the Landau fan diagram in different pressures, derived from the second derivative of the resistivity $\rho_{xx}$  with respect to the magnetic field squared, $d^2\rho_{xx}/dB^2$, as presented in Fig. \ref{s-figure4}(f)-(j). At zero pressure, the frequencies of the quantum oscillations, obtained from the slope of Landau fan diagram, are in good agreement with the one reported for the hole and electron pockets, 4.6 T and 6.5 T \cite{Zhu2010SM} respectively. Likewise, under pressure, the frequencies of quantum oscillations are also in good agreement with early early reports \cite{Brandt1980SM}. 

%\textcolor{red}{I don't understand the label of the frequency : at zero pressure $f_e$=13 (??) it should be smaller. For the hole it should be 4.5 T, and 6.5 T for the electrons. $f_h$ is the first peak, $f_e$ the second one and the third one is either 2 $f_e$ or $f_e$+$f_h$. Then for the other pressure it is not easy to follow and the data are not that nice. It seems that the background substraction is not good as you have a peak close to zero. REPLY: The measurements at low-field are not good enough to do FFT analysis. We now identify the Frequencies of quantum oscillations by directly labelling their peaks and found the consistence with previous reports.}

%. These SdH frequencies($F$) of the holes and electrons, marked in the figures, are closely related with the orthogonal extremal cross-sectional area($S$) of the Fermi pockets. 
Quantum oscillations allows also to extract the effective mass of electrons and holes by fitting the temperature dependence of the oscillation amplitude with the thermal damping term $\Delta R_{xx} = (14.69m^*T/B)/{\rm{sinh}}(14.69m^*T/B)$ by using the Lifshitz-Kosevich (L-K) formula \cite{Ramanayaka2010SM}, where $m^*$ is the effective cyclotron mass. The deduced pressure dependence of the frequencies and masses are shown in Fig. 4 (a) and (b) of the main manuscript.

%Although there is a large background, SdH oscillations are clearly visible in the second derivative  of $\rho_{xx}$ (See the data in the supplement).
\paragraph{Hall effect}

As the pressure increases the carrier density increases according to quantum oscillations study. It can also be verified by fitting the Hall effect in low field with the two-band model:

\begin{equation}
	\small{\rho_{xy}(B)=\frac{B}{e}\frac{(n_h\mu_h^2-n_e\mu_e^2)+(n_h-n_e)(\mu_e\mu_hB)^2}{(n_h\mu_h+n_e\mu_e)^2+(n_h-n_e)^2(\mu_e\mu_hB)^2}} \label{Rxy}
\end{equation}
where $n_{e,h}$ and $\mu_{e,h}$ are the electron and hole carrier density and mobility. Fig. \ref{s-figure5}(e) shows the results of the fit of $\rho_{xy}$ using Eq. \ref{Rxy}. Fig. \ref{s-figure5}(c) and (d) shows the deduced pressure dependence of $n_{e,h}$ and $\mu_{e,h}$. At ambient pressure $n_e=2.95 \times 10^{18}$ cm$^{-3}$ and $n_h=2.8 \times 10^{18}$ cm$^{-3}$, which are consistent with the previous reports \cite{Brandt1988SM,Fauque2016SM,Chapter1988SM}. The values of  $n_{e,h}$ and $\mu_{e,h}$ are also in agreement with the residual resistivity at zero magnetic field $\rho_{xx}(0)=\frac{1}{e}\frac{1}{n_e\mu_e+n_h\mu_h} \label{Rxx}$. The observed decrease of the mobility of carriers with pressure generally also agrees with the increase in Dingle temperature in graphite under pressure, see \cite{Brandt1980SM,Chapter1988SM}.

\paragraph{Pressure dependence of $v_{F,\parallel}$}
%Fig.\ref{s-figure5}(c) and (d) show the pressure dependence of carrier density and mobility which are increasing with pressure. 

%\textcolor{red}{I think there is a pb here. First after reading this paragraph I don't see how you arrive to the last equation between $m_{\parallel}$ and $m_{\perp}$. Second I guess you assume a parabolic dispersion. However it is not the case, see the last section. Then I am not sure how reliable is the pressure dependence of $v_{F,\parallel}$. REPLY: Yes, you are right. That's because we don't know $m^\ast_{\parallel}(0)$. We could not obtain the $v_{F,\parallel}$.}
The combination of the pressure dependence of the quantum oscillations, the in-planes masses and the carrier densities allows us to estimate the change of the Fermi velocity along the magnetic field ($v_{F,\parallel}$) with the pressure. 

\begin{figure}[htp]
	\includegraphics[width=8cm]{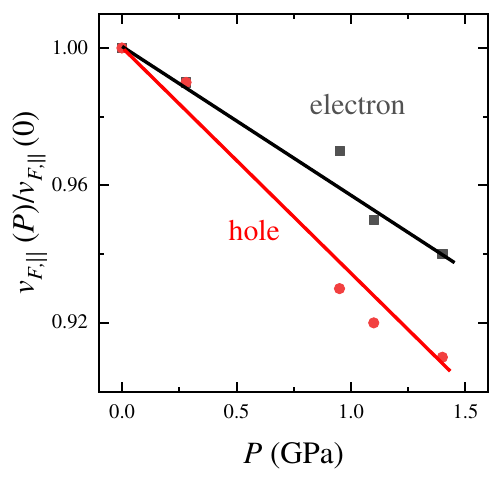}
	\caption{ %\textbf{Crystal structure, resistivity }
		\textbf{Pressure dependence of the Fermi velocity}  $v_{F,\parallel}=\frac{\hbar k_{F,\parallel}}{m^\ast_\parallel}$ of electrons (black) and holes (red).   }
	\label{s-v_F}
\end{figure}

Frequencies of quantum oscillations allow an accurate determination of ($k_{F,\perp}$) while the carrier density  $n(P)=\bar{k}^3_{F}(P)/3\pi^2$ where $\bar{k}_{F}(P)=(k^2_{F,\perp}(P)k_{F,\parallel}(P))^{1/3}$ which allow to determine $k_{F,\parallel}(P)$, the Fermi momentum along the $c$-axis. 
Assuming that the rate of change in mass under pressure remains constant but should reverse its direction between the plane and the $c$-axis ($\frac{m^\ast_{\parallel}(P)}{m^\ast_{\parallel}(0)}=\frac{m^\ast_{\perp}(0)}{m^\ast_{\perp}(P)}$), we can then estimate $v_{F,\parallel}(P)=\frac{\hbar k_{F,\parallel}(P)}{m_{\parallel}(P)}$.  Fig. \ref{s-v_F} shows the pressure dependence of $v_{F,\parallel}$ for the hole and electron.  They only decrease by about less than 10$\%$.

%\textcolor{red}{This last paragraph is not clear. REPLY: since we don't know the $m^\ast_{\parallel}(p)$, we rewrite this sentence:}
%Since we don't know how the $m^\ast_{\parallel}(p)$ evolves, we assume $\frac{m^\ast_{\parallel}(P)}{m^\ast_{\parallel}(0)}=\frac{m^\ast_{\perp}(0)}{m^\ast_{\perp}(P)}$, as shown in the in the Table \ref{mz}. The $v_{F,\parallel}(P)$ only changes by 9\% up to 1.4 Gpa, as shown in the Fig. \ref{s-v_F}.

\begin{table*}
	\centering
	\begin{tabular}{c|c|c|c|c|c|c|c|c}
		\hline
		Pressure (GPa)& $k_{F,h,\perp}(nm^{-1})$ & $k_{F,e,\perp}(nm^{-1})$ & $m^\ast_{h,\perp}(m_0)$ & $m^\ast_{e,\perp}(m_0)$ & $k_{F,h,\parallel}(nm^{-1})$ & $k_{F,e,\parallel}(nm^{-1})$ & $m^\ast_{h,\parallel}(m_0)$ & $m^\ast_{e,\parallel}(m_0)$ \\ \hline
		
		0  & 0.12 & 0.143  & 0.034  & 0.062  & 5.8  & 4.35  & 10 & 10  \\
		
		0.28  & 0.124 & 0.151 & 0.035  & 0.065 &  5.6 & 4.15  & 9.7 &  9.54 \\
		
		0.95  & 0.156 & 0.176 & 0.041  & 0.073 &  4.4 & 3.6  & 8.3 & 8.5  \\
		
		1.1  & 0.161 & 0.182& 0.043  & 0.076 &  4.2 & 3.4  & 7.92 &  8.16 \\
		
		1.4  & 0.168 & 0.196 & 0.045  & 0.08 &  4
		& 3.2  & 7.56 & 7.76  \\

		\hline
	\end{tabular}
	\caption{The mass and the Fermi wave-vector of electrons and holes at different pressure. }
	\label{mz}
\end{table*}

\section{Comparison of the SWM-model with the quantum oscillations frequencies}

%\textcolor{red}{Brandt \textit{et al.} and Itskevich \textit{et al.} have considered two possible approximations for the connection between the logarithmic derivatives of the parameters $\gamma_i(i=0-5)$, $\Delta$ and $\varepsilon_F$ with respect to pressure to deduce the $\partial ln|\gamma_2|/\partial P$. Brandt \textit{et al.} shows $\frac{\partial ln\gamma_1}{\partial P}\ne \frac{\partial ln\gamma_4}{\partial P}$ and  should be more accurate. We used the method of Brandt \textit{et al.} in the main text. Here is a detailed discussion: }

\begin{figure}[htp]
	\includegraphics[width=8cm]{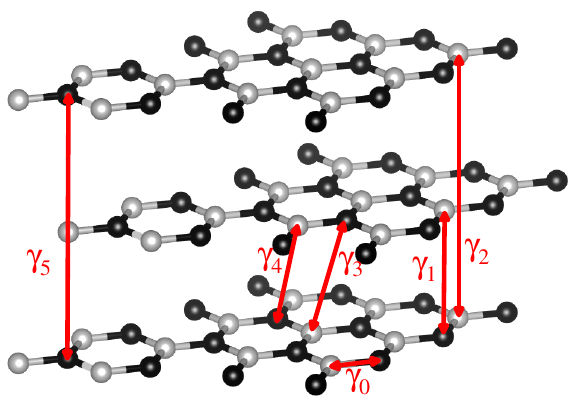}
	\caption{ 
		\textbf{stacked graphite}  The interaction between the carbon atoms are described by so-called the values of the parameters $\gamma_i(i=0-5)$.   }
	\label{s-gamma}
\end{figure}

The effective masses ($m^\ast$) and the extremal cross sections of the Fermi surface in the plane perpendicular to the $c$-axis are respectively described by the following formulas in the Slonczewski-Weiss-McClure (SWM) model\cite{Brandt1980SM}:

\begin{equation}
	m_e^\ast(\Psi)=\frac{4}{3}(\frac{\hbar}{a_0})^2\frac{\gamma_1}{\gamma_0^2}cos\Psi/(1+\frac{4\gamma_4}{\gamma_0}cos\Psi)     \label{me}
\end{equation}

\begin{equation}
	m_h^\ast(\Psi)=\frac{4}{3}(\frac{\hbar}{a_0})^2\frac{\gamma_1}{\gamma_0^2}cos\Psi/(1-\frac{4\gamma_4}{\gamma_0}cos\Psi)     \label{mh}
\end{equation}

\begin{equation}
	S_e=\frac{3\pi \hbar^2}{4\gamma_0^2a_0^2}\frac{2\gamma_2-\varepsilon_F}{(1+2\gamma_4/\gamma_0)^2}(\Delta-2\gamma_1+2\gamma_5+\varepsilon_F)     \label{Se}
\end{equation}

\begin{gather}
	S_h=\frac{3\pi \hbar^2}{4\gamma_0^2a_0^2}\frac{2\gamma_2cos^2\Psi_0-\varepsilon_F}{(1-2\gamma_4cos\Psi_0/\gamma_0)^2} \notag \\
	(\Delta+2\gamma_1cos\Psi_0+2\gamma_5cos^2\Psi_0-\varepsilon_F)   \label{Sh}
\end{gather}

\begin{equation}
	S_m=\frac{3\pi \hbar^2}{4\gamma_0^2a_0^2} \varepsilon_F(\varepsilon_F-\Delta)   \label{Sm}
\end{equation}

where $\Psi=k_zc_0/2$, $c_0=2c=6.7\mathring{A}$,$c$ is the inter-plane distance, $a_0=2a=2.462\mathring{A}$, $m_e^\ast$ ($m_h^\ast$) is the effective mass of the electron (hole). $S_e$ ($S_h$) is the extremal cross sections of the electron (hole). $\varepsilon_F$ is the Fermi energy, $\Delta=\gamma_6$, $cos\Psi_0\approx|\varepsilon_F/6\gamma_2|^{1/2}$. $S_m$ is the maximum section of the Fermi surface of the electron. The values of the parameters $\gamma_i(i=0-5)$ (see the Fig. \ref{s-gamma}), $\Delta$ and $\varepsilon_F$, (in eV) are given in Table \ref{gamma}.

\begin{table*}
	\centering
	\begin{tabular}{c|c|c|c|c|c|c|c|c}
		\hline
		$\gamma_0$ & $\gamma_1$ & $\gamma_2$ & $\gamma_3$ & $\gamma_4$ & $\gamma_5$ & $\Delta$ & $\varepsilon_F$ & Ref. \\ \hline
		3.2  & 0.397 & -0.0202  & 0.29  & 0.132  & 0.0098  & 0.0221  & -0.0223 & \cite{Brandt1980SM}  \\
		
		2.85  & 0.3 & -0.2 & 0  & 0 &  0 & 0.006  & -0.026 & \cite{Anderson1967SM}  \\
		
		%3.18  & 0.4 & -0.207 & 0.3  & 0.18 &  -0.006 & 0.005  & -0.024 & \cite{Johnson1973}  \\
		
		\hline
	\end{tabular}
	\caption{Band parameter sets for the SWM model.}
	\label{gamma}
\end{table*}

\begin{table*}
	\centering
	\begin{tabular}{c|c|c|c|c}
		\hline
		& $d{\rm{ln}}S_e/dP$ & $d{\rm{ln}}S_h/dP$ & $d{\rm{ln}}m^\ast_e/dP$ & $d{\rm{ln}}m^\ast_h/dP$  \\ \hline
		Itskevich \textit{et al.}\cite{Itskevich1967SM}  & 0.39 &   &   &    \\
		Anderson \textit{et al.}\cite{Anderson1967SM}  & $0.34\pm0.06$ & $0.4\pm0.04$ &   &    \\
		Mendez \textit{et al.}\cite{Mendez1980SM}  &   &   &   &    \\
		Brandt \textit{et al.}\cite{Brandt1980SM}  & $0.468\pm0.01$ & $0.485\pm0.01$ & $0.17\pm0.03$ & $0.24\pm0.05$  \\
		present work  & $0.467\pm0.02$ & $0.47\pm0.03$ & $0.182\pm0.026$ & $0.2\pm0.06$  \\
		\hline
	\end{tabular}
	\caption{Comparison of the logarithmic derivatives of the SWM-model parameter, the extremal cross sections and of the effective mass with pressure (in GPa$^{-1}$) in early works  \cite{Itskevich1967SM,Anderson1967SM,Mendez1980SM,Brandt1980SM} and in this work.}
	\label{comparison}
\end{table*}

%\textcolor{red}{What it is the point of this section ? I don't find it very useful. May be we should remove it. REPLY: There are two methods to deduce the $\partial ln|\gamma_2|/\partial P$. We tried to clarify this. We have added several sentences in the beginning of this section. }

Next, we calculate their pressure variation by taking the logarithmic derivatives of \ref{Sh} and \ref{Se} :

\begin{gather}
	2.232\frac{\partial ln|\gamma_2|}{\partial P}-1.23\frac{\partial ln|\varepsilon_F|}{\partial P}-
	0.03\frac{\partial ln|\Delta|}{\partial P}+1.03\frac{\partial ln\gamma_1}{\partial P} \notag \\
	-0.026\frac{\partial ln|\gamma_5|}{\partial P}-0.152\frac{\partial ln\gamma_4}{\partial P}=\frac{\partial lnS_e}{\partial P}   \label{dSe}
\end{gather}

\begin{gather}
	-0.5\frac{\partial ln|\gamma_2|}{\partial P}+1.56\frac{\partial ln|\varepsilon_F|}{\partial P}+0.057\frac{\partial ln|\Delta|}{\partial P}+0.876\frac{\partial ln\gamma_1}{\partial P} \notag \\
	+0.01\frac{\partial ln|\gamma_5|}{\partial P}+0.07\frac{\partial ln\gamma_4}{\partial P}=\frac{\partial lnS_h}{\partial P}   \label{dSh}
\end{gather}

%Itskevich and Anderson \textit{et al.} have considered the connection between the logarithmic derivatives of the parameters with respect to pressure \cite{Itskevich1967SM,Anderson1967SM}:

%\begin{equation}
%\frac{\partial ln|\gamma_2|}{\partial P}=\frac{\partial ln|\gamma_5|}{\partial P}=2\frac{\partial ln\gamma_1}{\partial P}=
%2\frac{\partial ln\gamma_3}{\partial P}=2\frac{\partial %ln\gamma_4}{\partial P}     \label{Anderson}
%\end{equation}

%Anderson \textit{et al.} obtained $a=0.24\pm0.04$ GPa$^{-1}$ from the extremal cross sections of the de Haas-van Alphen effect (dHvA). But from Eqs.(\ref{me})(\ref{mh}) and from the $m^*_e(P)$ and $m^*_h(P)$ dependencies obtained by Brandt \textit{et al.} it follows that $\frac{\partial ln\gamma_1}{\partial P}\ne \frac{\partial ln\gamma_4}{\partial P}$\cite{Brandt1980SM,Iye1990SM}:

%\begin{gather}
%\frac{\partial ln\gamma_1}{\partial P} 
%-0.066\frac{\partial ln\gamma_4}{\partial P}=\frac{\partial %lnm^\ast_e}{\partial P}   \label{dme}
%\end{gather}

%\begin{gather}
%\frac{\partial ln\gamma_1}{\partial P} 
%+0.076\frac{\partial ln\gamma_4}{\partial P}=\frac{\partial lnm^\ast_h}{\partial P}   \label{dmh}
%\end{gather}

%Thus they have therefore used another approximation in the calculations:

Following the same approximation as Brandt et al. \cite{Brandt1980SM}: 
\begin{equation}
	\frac{\partial ln|\gamma_2|}{\partial P}=\frac{\partial ln|\gamma_5|}{\partial P}=2\frac{\partial ln\gamma_1}{\partial P},  
	\frac{\partial ln\gamma_3}{\partial P}=\frac{\partial ln\gamma_4}{\partial P}     \label{Dillon}
\end{equation}

we can determine the pressure dependence of $m_e^\ast$ and $m_h^\ast$. The deduced pressure dependence  and of the extremal cross section are shown in Table \ref{comparison}. They are all in good agreement with early works \cite{Anderson1967SM,Brandt1980SM,Mendez1980SM}.

\section{The pressure dependence of DOS}

Along the $c$-axis, the DOS for the lowest Landau levels  follows: $E(k_z)$=-2$\gamma_2 \sin^2(\frac{c_0k_z}{2})$ where $\gamma_2$ is the interlayer hopping parameter between the two sub-lattices A and B and $c_0$=6.7$\AA$\cite{iye1982SM}. It follows that: $\frac{dE}{dk_z}$=-4$\pi\gamma_2 \sin{\pi \xi}\cos{\pi \xi}$. Close to the transition the band along the $c$-axis is close to half-filling or so, $k_z \approx \frac{\pi}{2c_0}$ which gives at first order that:  $\frac{dE}{dk_z}$=-$\pi\gamma_2$ which is a constant. This result reflects the fact that far from the extrema of the Landau level, the dispersion of the band along $k_z$ is almost linear. The DOS is thus constant and only depends of $\gamma_2$. Therefore $B^\ast$, like  $T^\ast$, scales as $\gamma_2$ and thus with $E_F$. Note that this result is different in the case of a parabolic dispersion along $k_z$, where the DOS scales with $\sqrt{E_F}$ and not $E_F$. In this case we would have expect a distinct pressure dependence for $T^\ast$ and $B^\ast$.

%For a parabolic dispersion along $k_z$, the DOS, inversely proportional to $B^*$, is expected to scale $\sqrt{E_F}$. But it is not the case of graphite, according to SWM model\cite{iye1982SM}. 

\section{The binding energy of exciton}

The exciton binding energy is given by $E_{\rm{B}}= (\mu/m_0)(1/\epsilon^2)R_{\rm{y}}$ \cite{Jerome1967SM,Halperin1968SM}, where $\mu$, $m_0$, $\epsilon$ and $R_{\rm{y}}$ are the exciton mass, the free electron mass, the dielectric constant and the Rydberg energy. Let's estimate its value in the case of graphite. 
By taking $m_\parallel$ = 5$m_0$\cite{Yaguchi2009SM}, $m_\perp$ = 0.25$m_0$\cite{Wang2020SM}, $\epsilon_\parallel=1.8225$\cite{Reyes2005SM,Palik1998SM} and the $\epsilon_\perp=10$\cite{1984IyeSM}, we found that, at zero pressure, $E_{\rm{B,\parallel}}=20.5 $eV, $E_{\rm{B,\perp}}=34 $meV. This exciton binding energy, is at least an order of magnitude larger than 10 K. Therefore, warming above 10 K destroys the order by crossing the Bose temperature of excitons (and not by unbinding the excitons).

\end{document}